# Title: Direct Measurement of Room Temperature Non-diffusive Thermal Transport Over Micron Distances in a Silicon Membrane

**Authors:** Jeremy A. Johnson[1]*†, A. A. Maznev[1]*, John Cuffe[2,3], Jeffrey K. Eliason[1], Austin J. Minnich[4]‡, Timothy Kehoe[2], Clivia M. Sotomayor Torres[2,5,6], Gang Chen[4], Keith A. Nelson[1].

**Affiliations:**

[1]Department of Chemistry, Massachusetts Institute of Technology, Cambridge, Massachusetts 02139, USA.

[2]Catalan Institute of Nanotechnology, Campus de Bellaterra, Edifici CM7, ES 08192, Barcelona, Spain.

[3]Department of Physics, Tyndall National Institute, University College Cork, Ireland.

[4]Department of Mechanical Engineering, Massachusetts Institute of Technology, Cambridge, Massachusetts 02139, USA.

[5]Catalan Institute for Research and Advanced Studies ICREA, 08010 Barcelona, Spain.

[6]Department of Physics, Universitat Autonoma de Barcelona, 08193 Bellaterra (Barcelona), Spain.

*Correspondence to: jeremy.johnson@psi.ch, maznev@mit.edu

†Current address: Paul Scherrer Institut, Villigen, Switzerland

‡Current address: California Institute of Technology, Pasadena, California

**Abstract**: The "textbook" phonon mean free path (MFP) of heat carrying phonons in silicon at room temperature is ~40 nm. However, a large contribution to the thermal conductivity comes from low-frequency phonons with much longer MFPs. We present a simple experiment demonstrating that room temperature thermal transport in Si significantly deviates from the diffusion model already at micron distances. Absorption of crossed laser pulses in a freestanding silicon membrane sets up a sinusoidal temperature profile that is monitored via diffraction of a probe laser beam. By changing the period of the thermal grating we vary the heat transport distance within the range ~1-10 μm. At small distances, we observe a reduction in the effective thermal conductivity indicating a transition from the diffusive to the ballistic transport regime for the low-frequency part of the phonon spectrum.

**Main Text:** The study of thermal transport at microscopic distances (*1-8*) is largely stimulated by practical needs such as thermal management of microelectronic devices (*2*), but also poses a number of fundamental physics problems. In dielectrics and semiconductors heat is carried predominantly by phonons, and the relationship between the phonon mean free path (MFP) and a characteristic length scale determines whether the thermal transport is diffusive or ballistic. At cryogenic temperatures phonon MFPs are relatively long and ballistic phonon propagation over macroscopic distances has been studied extensively (*9*). At room temperature, on the other hand, the majority of phonons have MFPs in the nanometer range. The often cited "textbook value" of the phonon MFP in Si at 273K based on a simple kinetic theory (*10*) is 43 nm, with even shorter MFPs listed for most other materials. According to this simplistic view one would not expect deviations from the classical thermal diffusion model at distances significantly exceeding 40 nm.

However, a growing body of experimental and theoretical studies has been indicating a large role of low-frequency phonons with MFPs much longer than tens of nanometers. Revising the "effective" room temperature phonon MFP in Si upwards to 260-300 nm has been suggested for the analysis of thermal transport in thin films (*11*) and superlattices (*12*). Recent measurements in Si have indicated non-diffusive transport on the tens of microns distance scale at temperatures 20-100 K (*7*). Still, it has been widely held that at room temperature heat transport in Si on the ~1 µm scale is consistent with diffusion theory (*1*).

On the theoretical side, first principles calculations of lattice thermal conductivity and phonon MFPs have emerged in recent years (*8,13-16*). Although quantitative discrepancies between different models still persist, they invariably point to a large contribution of low-frequency phonons to heat transport. For example, simulations by Henry and Chen (*13*) have indicated that phonons with MFP exceeding 1 µm contribute almost 40% to room temperature thermal conductivity of Si.

Measuring non-diffusive thermal transport at small distances in a configuration that can be quantitatively compared to theoretical models has been a challenge for experimentalists. Theoreticians favor the model of heat transport through a slab of material between two black body walls (*8,17*) which is all but impossible to realize in experiment. Just to mention one difficulty, any real interface between two materials involves thermal boundary resistance, which by itself presents a long-standing problem in nanoscale thermal transport (*1,18*). For a persuasive demonstration and to enable theoretical analysis beyond the diffusion model, an experiment should preferably (i) avoid interfaces, (ii) ensure one dimensional thermal transport, and (iii) clearly define the distance of the heat transfer and provide a way to vary this distance in a controllable manner. Experiments revealing non-diffusive transport on sub-micron length scales (*3-5*) were done with more complicated configurations involving heat transport from an irradiated film into a substrate, with the effective heat transfer distance in the substrate only indirectly inferred.

A method satisfying the above requirements has in fact been well known under the name laser-induced transient thermal gratings (*19,20*). In this method, two short laser pulses are crossed in a sample resulting in an interference pattern with period $L$ defined by the angle between the beams. Absorption of laser light leads to a spatially periodic temperature profile, and the decay of this temperature grating by thermal transport is monitored via diffraction of a probe laser beam. The heat transport from grating peaks to nulls does not involve any interfaces and the distance scale is controlled by the period of the optical interference pattern. An additional advantage of the method is a spatially sinusoidal temperature profile facilitating theoretical treatment.

In this report, we present transient thermal grating measurements of in-plane heat transport in freestanding silicon membranes. By varying the grating period we are able to directly measure the effect of the heat transfer distance on thermal transport (*21*).

The freestanding silicon membranes were fabricated by backside etching of silicon on insulator (SOI) wafers. In this process, the underlying Si substrate and buried oxide layer are removed through a combination of dry and wet etching techniques to leave a top layer of suspended silicon as shown in Fig. 1A (also see (*22*)). Measurements were conducted on two 400 nm thick membranes (membranes 1 and 2) with 400×400 µm² freestanding area fabricated on the same SOI wafer.

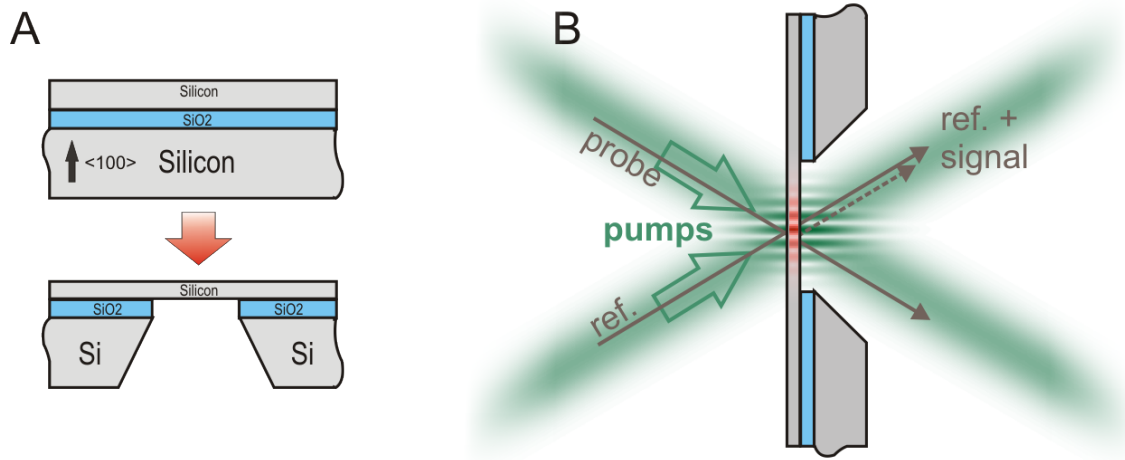

**Fig. 1**. Schematics of the sample and the experiment. A. Freestanding Si membranes are fabricated from SOI wafers by backside etching. B. Pump pulses are crossed in the silicon membrane, generating the transient thermal grating monitored via diffraction of the probe beam. Diffracted probe light is combined with a reference beam and directed to a fast detector.

Excitation laser pulses (wavelength $\lambda_e$ = 515 nm, pulse duration 60 ps) were crossed in the Si membrane with external angle $\theta_e$ as depicted in Fig. 1B. Interference between the two beams created a spatially periodic intensity and absorption pattern with interference fringe period $L=\lambda_e/2\sin(\theta_e/2)$. Above-bandgap photon absorption in the silicon membrane led to excitation of hot carriers, which promptly transferred energy to the lattice and relaxed to the bottom of the conduction band (*23*). Energy was deposited with a sinusoidal intensity profile resulting in a transient thermal "grating" with period $L$, i.e. with carrier population and induced temperature rise modulated as $(1 + \cos qx)$ where $q = 2\pi/L$ is the grating "wavevector" magnitude; excited carriers and heat subsequently diffused from grating peaks to nulls parallel to the surface. The membrane thickness was selected to be smaller than the ~1 µm absorption depth at the excitation wavelength to ensure one-dimensional in-plane heat transport.

Increased temperature and excited carriers induced changes in the complex transmittance, giving rise to time-dependent diffraction of a continuous wave probe beam (wavelength $\lambda_p$ = 532 nm). We used optical heterodyne detection whereby the diffracted signal was superposed with the local oscillator, or reference beam. Heterodyne detection not only increases the signal level but also yields a signal linear with respect to the material response that simplifies the interpretation and analysis of the data (*22*). A simple set-up using a diffraction grating to produce both excitation and probe-reference beam pairs ensures the precise overlap of the probe and reference beams as well as the stability of the heterodyne phase (*22,24*). The signal and reference beams were directed to a fast detector, whose output was recorded on an oscilloscope.

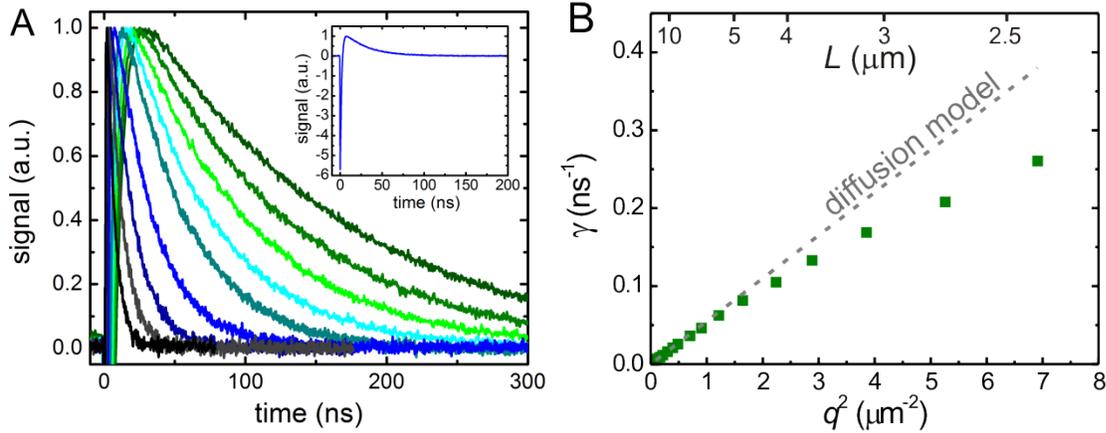

**Fig. 2**. Experimental data from membrane 1. A. Thermal decay traces for transient grating periods ranging from 3.2 to 18 μm. The decay time increases with the grating period. The inset shows the complete trace for the 7.5 μm period. B. Thermal grating decay rate versus the grating wavevector squared showing the departure from diffusive behavior. The dashed line representing the diffusion model was obtained by fitting the low-wavevector data in the range $L$ = 15-25 μm.

Data were collected at ~15 transient grating periods ranging from 2.4 to 25 μm in the two silicon membranes. Figure 2A shows traces collected from membrane 1 with transient grating periods from 3.2 to 18 μm. A complete waveform shown in the inset reveals a sharp negative peak due to electronic excitation. Fortunately, the ambipolar carrier diffusion coefficient in Si is about an order of magnitude greater than the thermal diffusivity (*25*). Thus electronic and thermal relaxations are well separated in the time domain: after the carrier population grating is washed out due to carrier diffusion, we are left with a purely thermal grating which decays more slowly due to phonon-mediated heat transport from grating peak to null. From the traces in Fig. 2A, we can see that the thermal decay becomes slower as the grating period increases; it takes longer for heat to move from grating peaks to nulls. According to the thermal diffusion equation, the temperature perturbation decays exponentially (*19*), $T(x,t) \propto \cos qx \exp(-\gamma t)$, with the decay rate $\gamma$ given by

$$\gamma = \alpha q^2 = kq^2/C, \tag{1}$$

where $\alpha$ is the thermal diffusivity, equal to the ratio of the thermal conductivity $k$ to the heat capacity per unit volume $C$.

We found that the thermal decay remains exponential within the whole range of grating periods (*22*). However, the decay rate deviates from the expected $q^2$ dependence as can be seen in Fig. 2B. This departure from diffusive behavior is even more apparent in Fig. 3A where we have plotted the effective thermal conductivity, obtained from the measured decay rate using Eq. (1), scaled by the bulk Si value, as a function of the grating period for the two membranes. At large grating periods, the thermal conductivity approaches a constant level, which is still significantly smaller than the bulk conductivity. It is well known that in-plane thermal conductivity of thin membranes is reduced due to scattering of phonons at the boundaries (*11,17*). However, as long as the diffusion model is valid, the thermal grating decay rate should vary as $q^2$, and the measured thermal conductivity value should remain independent of the grating period. We observed a significant further reduction in the measured thermal conductivity as the grating period was reduced below about 10 μm, clearly indicating a departure from diffusive thermal transport.

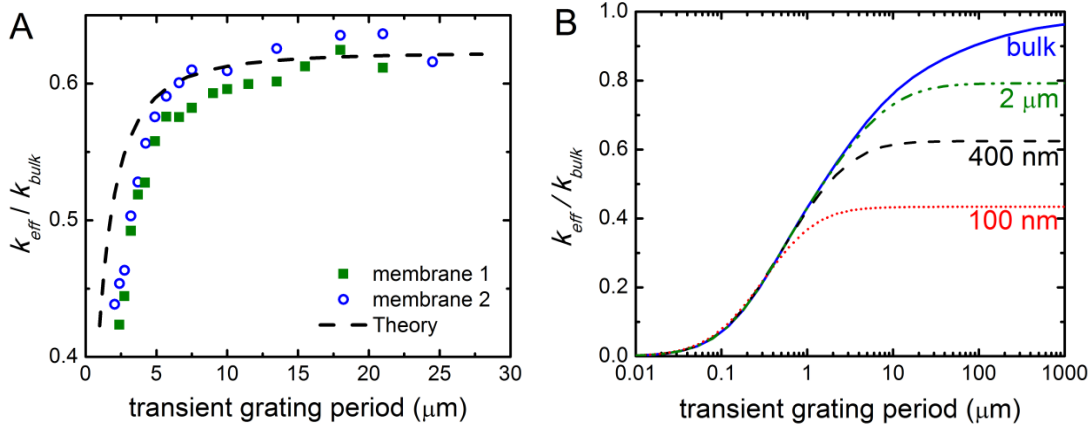

**Fig. 3**. A. The normalized effective thermal conductivity versus transient grating period compared with theory. B. The calculated effective thermal conductivity as a function of the grating period and the membrane thickness.

The decrease in the effective thermal conductivity is explained by the transition from diffusive to ballistic transport regime for low-frequency part of the phonon spectrum. There is no contradiction with the intuitive expectation that ballistic transport should be faster than diffusive: indeed, the grating decay is always faster at shorter periods as seen directly in Fig. 2A. However, the *increase in decay rate with wavevector* is slower than quadratic at short length scales since the traversal time of heat carried by ballistic phonons decreases linearly with distance, not quadratically as in the diffusive limit. In the relaxation time approximation which works well for Si above ~100K (*14*), thermal conductivity is given by the integral over the phonon spectrum,

$$k = \frac{1}{3}\int_0^{\omega_{max}} C_\omega v \Lambda d\omega , \qquad (2)$$

where $C_\omega$ is the differential frequency-dependent specific heat per unit volume, $v$ is the phonon group velocity, $\Lambda$ is the frequency-dependent MFP, and the summation over all phonon branches is implied. According to the Fourier law of heat conduction, the contribution of phonons at a given frequency to the heat flux is given by $Q_\omega = C_\omega v \Lambda \Delta T/3l$, where $l$ is the distance between the heat source and the heat sink and $\Delta T$ is the temperature difference. In this model the heat flux is supposed to increase indefinitely with increasing MFP, which cannot be true; obviously, it cannot exceed the purely ballistic black body radiation limit (*12*), $Q_{\omega\, bb} = C_\omega \Delta T/4$. Thus the contribution of ballistic phonons with $\Lambda \gg l$ to thermal transport will be suppressed *at least* by a factor of ¾$(l/\Lambda)$ compared to the predictions of the diffusion model. In the simplest approach, the contribution of all phonons with $\Lambda > l$ is simply disregarded, while for all phonons with $\Lambda < l$ the diffusion model is assumed to hold (*5,7*). In this case the "effective" thermal conductivity is found by simply cutting off the low frequency part of the integral in Eq. (2).

The simplicity of the transient grating geometry allowed us to develop a more rigorous theory based on the Boltzmann transport equation for phonons with MFP on the order of or larger than $l = L/2$ in combination with the diffusion equation for the "thermal reservoir" of high frequency phonons with $\Lambda<<l$ (*26*). We found that the grating decay remains exponential with the decay rate obtained by replacing the thermal conductivity in Eq. (1) by the effective conductivity,

$$k_{eff} = \frac{1}{3}\int_0^{\omega_{max}} AC_\omega v\Lambda d\omega$$

$$A(q\Lambda) = \frac{3}{q^2\Lambda^2}\left(1 - \frac{\arctan(q\Lambda)}{q\Lambda}\right) \quad ,$$

(3)

where the "correction factor" $A$ becomes unity in the diffusive limit $q\Lambda<<1$ and falls off as $(q\Lambda)^{-2}$ in the ballistic limit $q\Lambda>>1$. Unlike the simple "cut-off" model, Eq. (3) describes a smooth transition between diffusive and ballistic limits. The contribution of ballistic phonons to thermal transport is suppressed even more than according to the estimate based on the black body radiation limit because in the transient grating experiment the heat transport does not occur between black bodies. To the contrary, our heat "sources" and "sinks", i.e. maxima and minima of the thermal grating, become almost transparent for ballistic phonons in the limit $q\Lambda>>1$, which accounts for an additional factor of $\sim(q\Lambda)^{-1}$ in the ballistic phonon contribution to the heat flux.

In order to calculate the effective thermal conductivity according to Eq. (3), one needs to know the phonon density of states, group velocities and relaxation times for all phonon branches. For Si at room temperature, these quantities have been computed from first principles (*8,13-16*). We used the results of Ref. (*13*) presented in the form of thermal conductivity accumulation vs. MFP which is particularly convenient for our purposes (*22*). The calculated effective thermal conductivity for thermal grating relaxation in bulk Si is shown by the solid curve in Fig. 3B. The effective conductivity approaches the bulk value at large grating periods and decreases at small periods. The calculation is valid under the assumption (*26*) that diffuse phonons with $\Lambda<<L/2$ account for most of the specific heat, which holds well at $L>1$ μm.

In a thin membrane, the effective thermal conductivity is additionally reduced by boundary scattering. The classic formula for the effective MFP in a thin film was obtained (originally for electrons) by Fuchs (*27*). We combine the MFP reduction factor from the Fuchs-Sondheimer theory (*28*) with our Eq.(3) to estimate the combined effect of the heat transfer distance in the transient grating measurement and the boundary scattering in the membrane (*22*). In Fig. 3B, alongside the curve for bulk Si, we show the calculated results for three membrane thicknesses. In the large $L$ limit the effective thermal conductivity approaches a constant value determined by the membrane surface scattering. For thinner membranes, the onset of the non-diffusive effect is shifted towards shorter grating periods. As can be seen in Fig. 3A, the calculations for $d$=400 nm agree reasonably with the experiment given the uncertainties in the phonon MFP values obtained by different authors (*13,14*).

The fact that the deviations from the Fourier law in phonon mediated-thermal conductivity occur at much larger distances than previously thought should change the way we think of micro-scale thermal transport. One immediate implication is that accurate measurements of bulk thermal conductivity may be impossible on micron-sized samples. We have seen that the commonly cited textbook values of an "average" phonon MFP are of little relevance in analyzing the onset of size effects in thermal conductivity. Perhaps a more useful parameter would be the "median thermal conductivity MFP" $\Lambda_m$, such that phonons with $\Lambda > \Lambda_m$ contribute 50% to the bulk thermal conductivity. For Si at room temperature, calculations show this median MFP $\Lambda_m$ to be ~0.5-1 µm (*13,14,16*). The behavior of $\Lambda_m$ will be quite different from that of the "average" MFP. For example, impurity scattering makes all MFPs shorter; however, it affects primarily high-frequency phonons. Therefore $\Lambda_m$ may be in fact made larger by impurity scattering leading to larger size effects in semiconductor alloys compared to pure materials (*5*). For the same reason, we may expect larger size effects in thermal transport in natural diamond than in isotopically pure diamond contrary to what has been traditionally believed (*17*).

**Acknowledgments:** This work was supported as part of the S3TEC Energy Frontier Research Center funded by the U.S. Department of Energy, Office of Basic Energy Sciences under Award DE-SC0001299/DE-FG02-09ER46577 (experimental setup and data analysis). This work was also partially supported by projects: NANOPOWER, contract 256959; TAILPHOX, contract 233883; NANOFUNCTION, contract 257375; ACPHIN, contract FIS2009-150; AGAUR, 2009-SGR-150. The samples were fabricated using facilities from the "Integrated nano and microfabrication Clean Room" ICTS funded by MICINN.

**Supplementary Materials:**

**Si Membrane Fabrication**

Freestanding Si membranes were fabricated in nominally undoped, silicon-on-insulator (SOI) wafers using Si MEMS processing techniques (see Fig. S1). In this process, the underlying Si substrate and buried oxide layer are removed through a combination of dry and wet etching techniques to leave a top layer of suspended silicon.

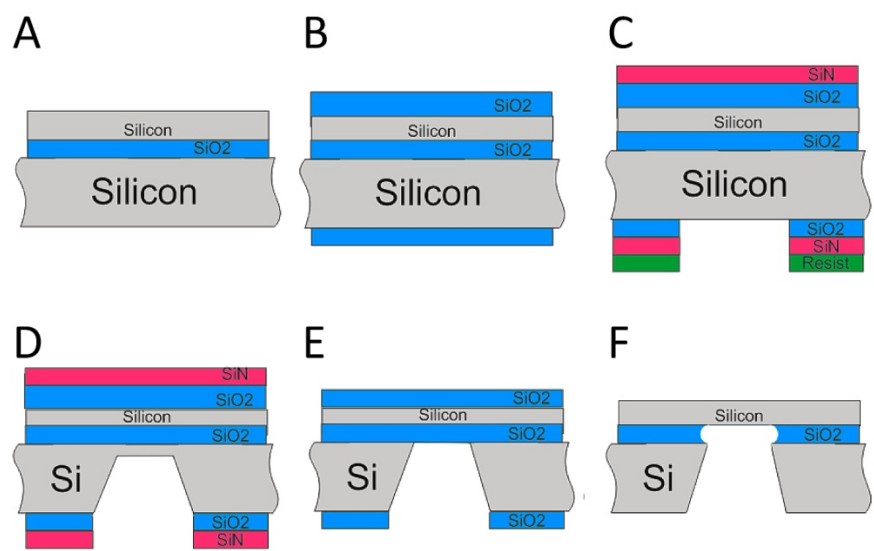

**Fig. S1**. Process flow for the fabrication of freestanding Si membranes: A. Original SOI wafer. B. Oxidation process to reduce thickness of Si layer. C. Deposition of $Si_3N_4$, photolithography of the etching window, including spin-coating, exposure, and development of the resist layer, and RIE of the $Si_3N_4$ and $SiO_2$ layers. D. Wet etching of the Si substrate using KOH, until approximately 1 μm of Si remains. E. Finishing of the etching of the substrate with TMAH, followed by removal of the $Si_3N_4$ by RIE. F. Final wet etching using HF to remove the top protective oxide and the buried oxide layers, releasing the freestanding Si membrane.

The initial SOI wafers were 625 μm thick in total, with a top Si layer thickness of ~1.5 μm and a ~1 μm buried oxide layer. The target thickness of the top Si layer was achieved through oxidation of the wafer; as the thermal oxide incorporates silicon during growth, the thickness of the top Si layer is reduced. For every unit thickness of Si consumed, 2.27 unit thickness of oxide is grown. The oxidation was performed at a temperature of 1100 °C in an atmosphere of water vapor in two steps to achieve fine control over the growth process, due to the large initial thickness of the top Si layer. Oxidation was continued until approximately 400 nm of Si remained on the top layer, and the thermal oxide was left as a protective layer, as illustrated in Fig. S1B, until the final stage of the process.

A silicon nitride layer was deposited to act as a mask during the subsequent wet etching of the Si substrate and the freestanding areas of the membranes were determined through photolithography on the backside of the wafer, involving spin-coating of a photoresist, exposure, and development. The remaining photoresist was then used as a mask for Reactive Ion Etching (RIE) to open etching windows in the $Si_3N_4$ and $SiO_2$ (see Fig. S1C).

A wet etching process with potassium hydroxide (KOH) and tetramethylammonium hydroxide (TMAH) was used to remove the Si substrate. The selectivity of TMAH is better when using a $Si_3N_4$ or $SiO_2$ mask for the etching of Si, though the etch rate is slower. The layer of $Si_3N_4$ was deposited to improve the etch selectivity compared to the $SiO_2$, and the Si substrate was etched with KOH until approximately 1 μm of Si substrate remained (Fig. S1D). The etching of the substrate was finished with TMAH, and the $Si_3N_4$ was then removed by RIE. (Fig. S1E). The etching occurs preferentially in the <100> direction with an etching angle of 54.7 degrees, thus the areas of the membranes are substantially smaller than the original patterns on the backside of the wafer. The reduction in length of one side can be calculated approximately by $x_f = x_i - 2d_{sub}$, where $d_{sub}$ is the thickness of the Si substrate, $x_i$ is the initial length on the backside of the wafer, and $x_f$ is the final length on the topside of the wafer. This relaxes the resolution requirements for the photolithography, and allows the photolithography masks to be produced from inexpensive, disposable acetate in place of quartz.

After the wet etching process of the Si substrate, the top, bottom, and buried oxides were removed by a wet etch of hydrofluoric acid (HF) to release the freestanding Si membranes (Fig. S1F). Measurements were conducted on two membranes with 400×400 μm² freestanding area and thicknesses, determined by optical reflectometry, of 400±10 nm.

**Experimental Details**

A short-pulsed excitation laser beam was derived through the frequency doubled output ($\lambda_e$ = 515 nm) of an amplified Yb:KGW laser system (HighQ femtoRegen, set to 1 kHz repetition-rate). Although the laser is designed to output pulses as short as 300 fs in duration, to avoid sample damage and unwanted nonlinear optical effects from high peak powers, we have bypassed the compressor to obtain ~60 ps pulses. As depicted in Fig. S2, the pump beam was split with a custom diffractive optic (a binary phase mask pattern) into two beams which were passed through a two-lens telescope (with 2:1 imaging by achromatic doublets) and were focused and crossed within the membrane, with 3.6 μJ per pulse and the spot size radius 300 μm at 1/e intensity level. The CW probe beam, derived from the output of a single-longitudinal-mode, intracavity frequency-doubled Nd:YAG laser at 532 nm, followed an almost identical optical path as depicted in Fig. S2 and was focused into a spot of 150 μm 1/e radius aligned with the center of the excitation spot. To reduce sample heating, an electro-optic modulator was used in conjunction with a delay generator to chop the probe beam into 64 μs rectangular pulses synchronized with the pump pulses. An absorptive neutral density filter was used to attenuate the 5.2 mW reference beam power by a factor of 1000 to avoid detector saturation. The heterodyne phase was controlled by small angle adjustments of a highly parallel fused silica plate placed in the probe beam path. The co-propagating reference and diffracted probe beams were directed to a Hamamatsu C5658 silicon avalanche photodiode with 1 GHz bandwidth and the signal traces were recorded on an oscilloscope with 4 GHz bandwidth. Traces of 40,000 averages were subsequently downloaded to the computer for data analysis. The temperature in the lab varied between 296 and 298 K.

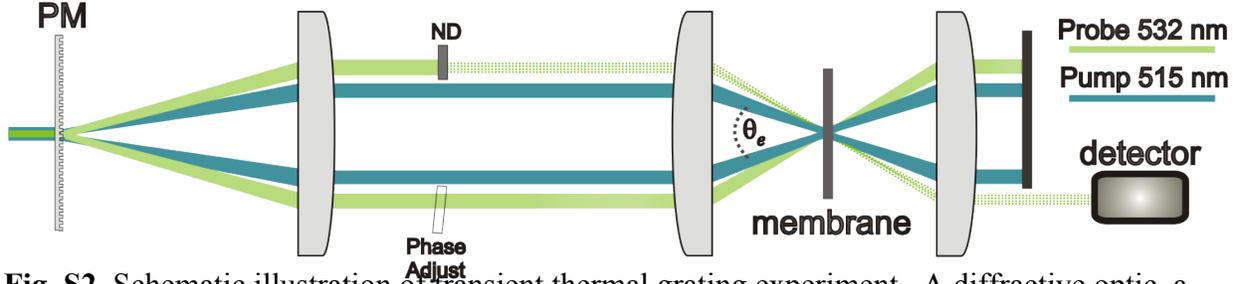

**Fig. S2**. Schematic illustration of transient thermal grating experiment. A diffractive optic, a binary phase-mask (PM), splits pump and probe into ±1 diffraction orders. Pump beams are crossed in the silicon membrane, generating the transient thermal grating. Diffracted probe light is combined with a reference beam attenuated by a neutral density filter (ND) and directed to a fast detector. The relative phase difference between probe and reference beams is controlled by adjusting the angle of a glass slide (Phase Adjust) in the probe beam path.

**The Probing Process and Heterodyne Detection**

The excitation processes and resulting material responses described above lead to time-dependent, spatially periodic changes in both the complex refractive index of silicon and the thickness of the membrane; the dynamics of these induced transient grating responses will be encoded in the diffracted probe light that is directed to a detector. The optical fields of the probe and reference beams incident on the sample are approximated, respectively, as plane waves

$$E_p = E_{0p} \exp\left( i\left(k_p^2 - q^2/4\right)^{1/2} z - i(q/2)x - i\omega_p t + i\phi_p \right) \quad \text{(S1)}$$

and

$$E_R = t_r E_{0p} \exp\left( i\left(k_p^2 - q^2/4\right)^{1/2} z + i(q/2)x - i\omega_p t + i\phi_R \right) \quad \text{(S2)}$$

where $E_{0p}$ is the incident probe amplitude, $t_r$ is the attenuation factor for the reference beam, $q$ is the transient grating wavevector, $k_p$ is the optical wavevector magnitude, $\omega_p$ is the optical frequency, and $\phi_p$ and $\phi_R$ are the phases of probe and reference beams respectively.

The diffracted field can be obtained by multiplying the input field by the complex transfer function (*30*) for an absorbing slab (*31*). The transfer function depends on the complex refractive index and the thickness, both of which are functions of temperature. Assuming the temperature grating to be a small perturbation, the transfer function can be represented as

$$t = t_0 [1 + a(T_u + T_g \cos(qx))], \quad \text{(S3)}$$

where the temperature perturbation is comprised of a spatially uniform component $T_u$ and the grating component $T_g$, and $a$ is the magnitude of the effect taking into account changes in the complex refractive index and sample thickness.

Assuming that the sample is located at $z = 0$, for the +1 diffraction order of the probe beam one obtains

$$E_{p(+1)} = \frac{1}{2} t_0 a T_g E_{0p} \exp\left( i\left(k_p^2 - q^2/4\right)^{1/2} z + i(q/2)x - i\omega_p t + i\phi_p \right) \quad \text{(S4)}$$

and for the zero order reference beam

$$E_{R(0)} = t_r t_0 (1 + aT_u) E_{0p} \exp\left( i\left(k_p^2 - q^2/4\right)^{1/2} z + i(q/2)x - i\omega_p t + i\phi_R \right) \quad \text{(S5)}$$

The two beams are collinear and their interference gives an intensity

$$I_s = \frac{1}{2} I_{0p} |t_0|^2 \left[ t_r^2 |1 + aT_u|^2 + |a|^2 T_g^2 + t_r T_g \left( ae^{i\phi} + a^* e^{-i\phi} \right) + 2t_r |a|^2 T_u T_g \cos\phi \right], \tag{S6}$$

where $I_{0p}$ is the intensity of the probe beam, $a^*$ is the complex conjugate of $a$, and $\phi = \phi_p - \phi_R$ is the heterodyne phase. The heterodyne phase is well defined and easily adjusted by rotating the thin glass plate in the probe beam path (Fig. S2). Neglecting terms quadratic with respect to the temperature perturbation, we get

$$I_s = \frac{1}{2} I_{0p} |t_0|^2 \left[ t_r^2 \left(1 + 2T_u \operatorname{Re} a \right) + 2t_r T_g \left( \operatorname{Re} a \cos\phi + \operatorname{Im} a \sin\phi \right) \right]. \tag{S7}$$

We collect the signal for two values of the heterodyne phase $\phi$ separated by $\pi$ and calculate the difference equal to

$$I_{diff} = t_r |t_0|^2 I_{0p} T_g \left( \operatorname{Re} a \cos\phi + \operatorname{Im} a \sin\phi \right). \tag{S8}$$

The dynamics of this difference signal are entirely determined by the amplitude of the temperature grating $T_g$. The heterodyne phase is adjusted to maximize the signal; the exact value of the phase does not affect the signal shape. Taking the difference of the signals measured at two opposite heterodyne phases also helps suppress spurious signals, such as Pockels' cell interference, that are not sensitive to the heterodyne phase.

If the coefficient $a$ is real, then we have an amplitude transient grating; if $a$ is imaginary, then it is a phase grating. The coefficient $a$ can be calculated if the temperature dependence of the refractive index, the photoelastic constants, and the thermal expansion coefficient are all known. For our purposes the exact value of $a$ is unimportant because it only affects the signal amplitude, while the dynamics of the thermal grating decay are extracted from the temporal profile of the signal. Schmotz *et al.* (*29*) reported the temperature-dependent transmissivity of a similar freestanding silicon membrane showing that small changes in temperature can lead to substantial changes in the transmission at all optical wavelengths, but only small changes in position of interference fringes, indicating that a periodic, transient change in temperature will result in a strong amplitude transient grating, potentially much greater in magnitude than a phase grating.

**Data Analysis**

To measure the decay rate of the thermal grating, one could exclude the initial fast electronic response and fit the rest of the signal to a single exponential decay. However, this could introduce a small systematic error depending on the position of the fit starting point. Because the electronic response is expected to follow a diffusion equation similar to the thermal response (*25*), we chose to account for the electronic response directly by fitting the signal to a bi-exponential form as illustrated in Fig. S3A for the signal waveform obtained at $L=10$ μm. Single-exponential components of the fit function corresponding to the electronic and thermal components of the signal are shown as well; one can see that the two components are well separated in time. Figure S3B shows good agreement between the data and bi-exponential fit for four representative grating periods (note the logarithmic time scale chosen to show both electronic and thermal dynamics at different grating periods). Oscillations present in the 2.4 μm signal are due to the detector response to the fast initial electronic signal.

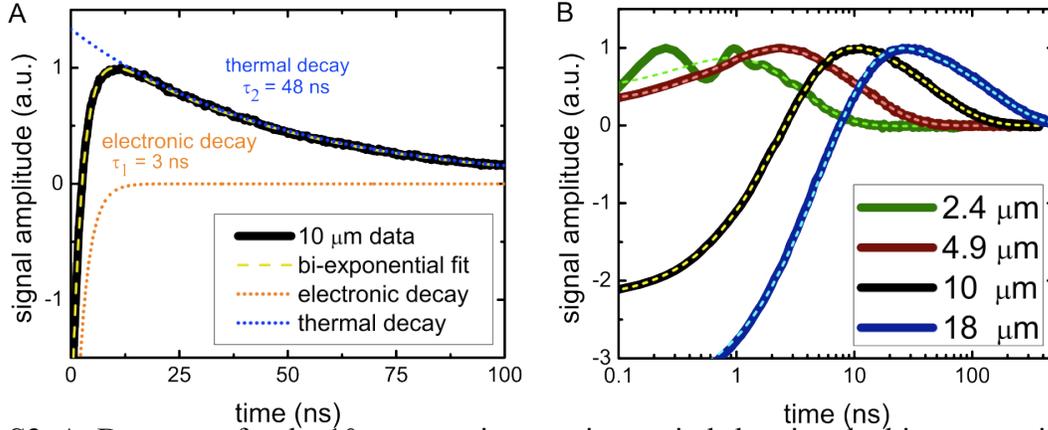

**Fig. S3**. A. Data trace for the 10 μm transient grating period showing the bi-exponential fit (dashed line) in tandem with fast electronic and slower thermal exponential decays (dotted lines). B. Full traces of 2.4, 4.9, 10, and 18 μm transient grating periods with accompanying bi-exponential fits (dashed lines).

### Details of the effective thermal conductivity calculations

By changing the integration variable in Eq. (2), thermal conductivity can be represented as an integral over MFP,

$$k = \int_0^\infty k_\Lambda d\Lambda ,\qquad(S9)$$

with the differential thermal conductivity $k_\Lambda = \frac{1}{3} C_\omega v \Lambda (d\Lambda/d\omega)^{-1}$ readily found from the thermal conductivity accumulation data with respect to MFP presented by Henry and Chen (*13*). The effective thermal conductivity in the transient grating geometry is found by multiplying the integrand by the correction factor,

$$k_{eff} = \int_0^\infty A(q\Lambda) k_\Lambda d\Lambda .\qquad(S10)$$

The above equation was used to calculate the solid curve in Fig. 3B.

For a thin membrane, we need to account for the effect of boundary scattering. In Fuchs-Sondheimer theory, the effective MFP reduced due to diffuse scattering at the surfaces of the membrane is given by (*28*)

$$\Lambda' = \Lambda \Phi\left(\frac{d}{\Lambda}\right)$$
$$\Phi(\chi) = 1 - \frac{3}{8\chi} + \frac{3}{2\chi}\int_1^\infty \left(\frac{1}{t^3} - \frac{1}{t^5}\right) e^{-\chi t} dt \qquad(S11)$$

where *d* is the membrane thickness. The correction factor Φ behaves similarly to the factor *A* in that it approaches unity in the limit of small MFP (Λ<<*d*) and drops off when the MFP exceeds *d*.

Equation (S11) can be modified to allow for partially specular reflections (*28*). However, using the specular reflection probability as a free parameter introduces considerable arbitrariness in the data analysis; moreover, the commonly used model with a constant "specularity parameter" independent of both frequency and incidence angle (*28*) is rather unphysical. Most studies of thermal conductivity in thin films use the "diffuse scattering" model, with satisfactory results (*11,32,33*).

To account for the combined effect of the heat transfer distance in the transient grating measurement and the boundary scattering in the membrane we take the MFP reduced by the boundary scattering from Eq. (S11) and plug it into Eq. (S10), which yields the following result:

$$k_{eff} = \int_0^\infty A(q\Lambda\Phi) k_\Lambda \Phi d\Lambda \ .  \tag{S12}$$

This equation, with $k_\Lambda$ from Henry and Chen (*13*), was used to produce the theoretical curves for thin membranes in Fig. 3. Equation (S12) is of course an approximation. For a rigorous analysis, one would need to solve the non-equilibrium thermal transport problem anew with appropriate boundary conditions at the surfaces of the membrane, which would be much harder than the analysis for an unbounded medium (*26*). Note that the wavevector dependence in Eq. (S12) is only present in the correction factor *A* resulting from non-diffusive transport.